\documentclass[preprint,onecolumn]{pnastwo}
\usepackage[dvips]{graphicx}
\usepackage{color}
\usepackage{times}

\usepackage{amssymb,amsfonts,amsmath,mathrsfs}

\begin{document}

\title{Electrostatics-driven shape transitions in soft shells}

\author{Vikram Jadhao\affil{1}{Departments of Materials Science and Engineering}, Creighton K. Thomas\affil{1}{},\and Monica Olvera de la Cruz\affil{1}{}\affil{2}{Chemistry}\affil{3}{Chemical and Biological Engineering}\affil{4}{Physics, Northwestern University, Evanston, IL 60208, USA}}


\maketitle

\begin{article}

\begin{abstract} 
Manipulating the shape of nanoscale objects in a controllable fashion is at the heart of designing materials that act as building blocks for self-assembly or serve as targeted drug delivery carriers. Inducing shape deformations by controlling external parameters is also an important way of designing biomimetic membranes.  In this article, we demonstrate that electrostatics can be used as a tool to manipulate the shape of soft, closed membranes by tuning environmental conditions such as the electrolyte concentration in the medium. Using a molecular-dynamics-based simulated annealing procedure, we investigate charged elastic shells that do not exchange material with their environment, such as elastic membranes formed in emulsions or synthetic nanocontainers. We find that by decreasing the salt concentration or increasing the total charge on the shell's surface, the spherical symmetry is broken, leading to the formation of ellipsoids, discs and bowls. Shape changes are accompanied with a 
significant lowering of the electrostatic energy and a rise in the surface area of the shell. To substantiate our simulation findings, we show analytically that a uniformly-charged disc has a lower Coulomb energy than a sphere of the same volume. Further, we test the robustness of our results by including the effects of charge renormalization in the analysis of the shape transitions and find the latter to be feasible for a wide range of shell volume fractions. 
\end{abstract}


\section{Significance}
Shape is a fundamental property of an object that influences its interaction with the environment and often determines the object's functional capabilities. Understanding how to generate and control shape by modifying the environmental conditions is of primary importance in designing systems that respond to external clues. We show here that electrostatic interactions can be used to change the equilibrium shape of soft, nanometer-sized shells. We find that a uniformly-charged, spherical shell undergoes shape changes, transforming into ellipsoids, discs, and bowls, as the electrolyte concentration in the environment is decreased. This electrostatics-based shape design mechanism, regulated by varying properties external to the shell, can be used to build efficient nanocontainers for various medical and technological applications. 
\\

\dropcap{B}iological matter in cells is often compartmentalized by elastic membranes that take various shapes such as blood cell membranes, organelles and viral capsids. These biomembranes are highly optimized to perform specific functions. A key focus of current biomedical technologies is to engineer synthetic materials that can match the performance and structural sophistication displayed by natural entities. Mimicking key physical features of biomembranes, including shape, size, and flexibility, is a crucial step towards the design of such synthetic biomaterials \cite{doshi}. Recent findings also indicate that the shape of a drug-carrier nanoparticle directly influences the amount and efficiency of drug delivery  \cite{discher1,mitragotri1,mitragotri2,loverde}. The shape and deformability of soft materials such as colloids, emulsions, hydrogels or micelles play an important role in determining their usefulness in various technological applications as well \cite{Cademartiri,suo,xia1,nanoshells}. 
For example, colloidal self-assembly is governed to a large extent by the shape of individual colloids \cite{Cademartiri, sacanna1, sacanna2}. 
Similarly, controlling the shape and size of reverse micelles is of key importance in their use as solvent extraction systems for removing rare-earth metals from aqueous solutions or as templates for nanoparticle synthesis \cite{lok,pileni,solvent_extraction,hoogerstraete}. 

Shape transformations in materials are engineered via chemically-induced modifications \cite{sacanna1,sacanna2} or using techniques such as photoswitching of membrane properties \cite{hamada} and controlled evaporation of the enclosed solvent \cite{quilliet}. However, generating desired material shapes with precision and manipulating them with relative ease at the nanoscale has been a challenge \cite{Cademartiri}. From the theoretical standpoint, much attention has been focused on finding the low-energy conformations of flexible materials, modeled often as soft elastic membranes, in the hope of suggesting superior experimental systems that can enable the design of nanostructures  \cite{vernizzi2,bowick2,mahadevan}. Examples include the exploration of shape transitions driven by topological defects \cite{seung-nelson,lidmar,siber} or compression \cite{gompper}, and the study of low-energy conformations of multicomponent shells \cite{vernizzi2,datta,rastko2,funkhouser}.

Changing the shape of an elastic shell entails bending and stretching it and the associated energy costs form the components of the elastic free energy of the shell \cite{landau_toe}. However, when the shell is charged, it is possible to compensate for the increase in elastic energy associated with the shape deformation if the latter is accompanied with a significant lowering of the electrostatic free energy \cite{vernizzi,rastko1,grohn,leung,vorobyov}. Previous studies on charged, soft membranes mainly focused on mapping a charged elastic shell to an uncharged elastic shell with charge-renormalized elastic parameters \cite{helfrich,duplantier,pincus,kim-sung,netz2,andelman}. In the case of charged nanoshells, electrostatic screening length is comparable to the shell dimensions and the surface charge density can assume high values. As a result, shell models where Coulomb interactions are included explicitly are needed \cite{vernizzi,rastko1}. Using such models, it has been shown that an ionic shell, where positive and negative charges populate the surface, lowers its energy by taking an icosahedral shape with the same surface area \cite{vernizzi}. In this work, we find that a uniformly-charged, spherical elastic shell, when constrained to maintain the enclosed volume, can lower its free energy by deforming into smooth structures such as ellipsoids, disks, and bowls (see Fig.~\ref{fig1}). We show that the transition to these nonspherical shapes can be driven by varying environmental properties such as the electrolyte concentration in the surrounding solvent.  

In order to include the non-linear coupling between the shape of the shell and its electrostatic response self-consistently, we study the soft, charged nanoshells numerically. We model the charged shell by a set of discrete points placed on a spherical membrane, forming a mesh consisting of vertices, edges and faces, recognizing that in the limit of large number of vertices, the discretized elastic membrane recovers the physics of the associated continuum model (see \emph{Materials and Methods} for details). The uniform surface charge density is simulated by assigning every vertex with the same charge. We work with elastic parameters such that the uncharged shell assumes a spherical shape at equilibrium. We allow only the deformations that preserve the shell's total volume, the latter being chosen to be that of the uncharged conformation. Our model is applicable to monolayers, such as emulsions or reverse micelles where nanodroplets of oil or water are surrounded by properly polymerized 
charged surfactant molecules, and also to incompressible bilayer systems and nanocontainers that do not exchange material with their environment. In the following sections, we provide evidence that this minimum model reproduces various shapes observed experimentally. Furthermore, we test the validity of this electrostatic model and associated simulation results by providing analytical solutions in limiting cases, namely by computing the electrostatic energy of oblate spheroidal shells and comparing it to that of a sphere of the same volume in salt-free conditions. Effects of ion condensation are then included via a two-state model to derive the renormalized charge on the spherical and spheroidal shells in order to test the robustness of our results.

Using the discretization of the continuum expression for the elastic energy introduced in Ref.~\cite{seung-nelson}, we write the free energy $\mathcal{F}$ associated with the discretized shell as
\begin{eqnarray}\label{eq:free_energy}
\mathcal{F}[\{\mathbf{r}_{i}\}] &=& \frac{\kappa}{2} \sum_{l\in \rm{E}} |\mathbf{n}_{l,1} - \mathbf{n}_{l,2}|^{2} + \frac{k}{2R^2} \sum_{l\in \rm{E}} \left(\left|\mathbf{r}_{l,1} - \mathbf{r}_{l,2}\right| - a_l\right)^{2} \nonumber\\
&& +\, \frac{l_{B} z^2}{2} \sum_{i,j\in \rm{V}} \frac{e^{-|\mathbf{r}_{i} - \mathbf{r}_{j}| /\lambda_{\rm D}}}{|\mathbf{r}_{i} - \mathbf{r}_{j}|},
\end{eqnarray}
where $\mathcal{F}$ is measured in units of $k_B T$. Here $T$ is the room temperature and $k_{B}$ is the Boltzmann constant. We make the free energy dimensionless by defining $\kappa = \tilde \kappa/k_B T$, where $\tilde \kappa$ is proportional to the bending rigidity $\kappa_{\textrm{b}}$ of the continuum model, and $k = \tilde k R^2 / k_B T$, with $\tilde k$ being proportional to the 2D Young's modulus $Y$ of the continuous elastic membrane, and $R$ is the spherical shell radius. We employ the dimensionless bending rigidity $\kappa$ and the spring constant $k$ as the scale for bending and stretching energies respectively. In Eq.~\eqref{eq:free_energy}, $\rm E$ and $\rm V$ denote the set of all edges and vertices respectively, and $\mathbf{r}_{i}$ is the position vector of the $i^{\rm th}$ vertex. The first term on the right-hand side is the bending energy with $\mathbf{n}_{l,1}$ and $\mathbf{n}_{l,2}$ being the normal vectors to the faces adjacent to edge $l$. The second term is the stretching energy with $\mathbf{r}_{l,1}$ and $\mathbf{r}_{l,2}$ being the position vectors of the vertices corresponding to the edge $l$, and $a_l$ is the rest length of edge $l$. 
The last term is the (dimensionless) electrostatic energy of the model membrane. We consider an aqueous environment inhabiting  electrolyte whose presence is taken into consideration implicitly, leading to screened Coulomb interactions between each vertex pair. Here, $l_{\rm B}$ denotes the Bjerrum length in water, $\lambda_{\rm D}$ is the Debye length and $z$ is a dimensionless charge associated with each vertex. We assume a uniform dielectric in order to simplify the computations, thus ignoring any induced charge effects.  

As is evident from Eq.~\eqref{eq:free_energy}, the free energy $\mathcal{F}$ is a function of the set of vertex position vectors $\{\mathbf{r}_{i}\}$ which also parametrizes the shape of the shell. The equilibrium shape of the shell is the one that corresponds to the minimum of $\mathcal{F}$ subject to constraint of fixed enclosed volume. We perform this constrained free-energy minimization using a molecular dynamics (MD) based simulated annealing procedure, details of which are provided in \emph{Materials and Methods}.  

\section{Results}
The uncharged elastic shell conformation in all our simulations is a sphere of radius $R \approx 10$ nm. We discretize the sphere with $N \approx 1000$ points, generating a nearly-uniform distribution with an average edge length of $a \approx 1$ nm. We fix the elastic spring constant $k = 100$ in all simulations. This value corresponds to $Y = 1$ $k_{B}T$ per $\textrm{nm}^{2}$ which, for the bending rigidities under investigation, leads to shells characterized by a F\"oppl-von K\'arm\'an number ($=YR^2 / \kappa$) in the range $10 - 100$. We consider a monovalent electrolyte with concentration $c$. The Debye length $\lambda_{\rm D}$ is known via the relation $\lambda_{\rm D} = 0.304/\sqrt{c}$ nm \cite{israel}. The concentration $c$ thus parametrizes the spatial range of Coulomb interactions. In our simulations, we tune $c$ such that this range varies from $\lambda_{D} < a$, in which case the shell mimics the behavior of an uncharged elastic shell, to $\lambda_{D} \sim R$, which corresponds to the case where most charges feel each other. 

\begin{figure}
\centerline{
\includegraphics[scale=0.1]{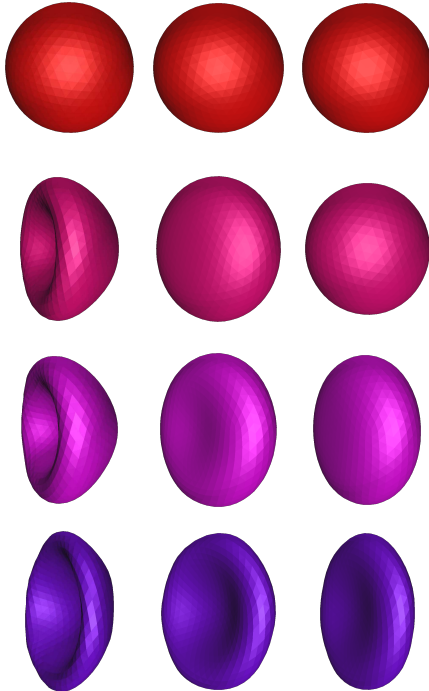}
}
\caption
{
\label{fig1}
Snapshots of minimum-energy conformations of charged elastic nanoshells for three different bending rigidities  $\kappa =$ 1, 5, 10 (columns, left to right). In each column the electrolyte concentration $c$ (M) decreases (top to bottom) as $c = 1, 0.1, 0.05, 0.005$. Different colors suggest different concentration values, with red being the highest $c$ under study and purple corresponding to the lowest $c$. As the concentration is lowered, the range of electrostatic interactions is increased, leading to the variation in the shape of the nanoshell. We find that for the concentration-range under investigation, softer shells tend to form bowl-like structures, while more rigid vesicles form ellipsoidal and disk-like shapes. All the above nanostructures have the same total surface charge and volume, fixed to values associated with the spherical conformation.
}
\end{figure}
In Fig.~\ref{fig1}, we show the change in the shape corresponding to the minimum of the shell free energy $\mathcal{F}$ as $c$ is varied. Here, we set $z = 0.6$, which is equivalent to a shell surface potential of $\approx 100$ mV. Each column represents the shapes obtained for a fixed value of $\kappa$, with the latter increasing from left to right assuming the values $\kappa =$ 1, 5, 10. Within each column, $c$ decreases  from top to bottom as $c =$ 1, 0.1, 0.05, 0.005 M. This range of concentration covers most biological and synthetic conditions. We see that the top row ($c = 1$ M) is comprised of spherical shapes. At $c = 1$ M, the screening length is very small ($\lambda_{\textrm{D}} < a$) and hence the electrostatic forces only come into effect at extremely short distances, resulting in a nearly vanishing contribution to the overall free energy. This leads to conformations that resemble the shape of the uncharged elastic shell which is spherical. However, as $c$ is lowered, transitions to a variety of nonspherical shapes are observed. 

In case of the most flexible charged shell (Fig.~\ref{fig1}, left column), increasing the range of the electrostatic interactions leads to the formation of concave structures, hereafter referred to as bowls. The opening of the bowl widens with decreasing $c$. For a shell with a higher bending rigidity (middle column), as $c$ is lowered, the shell first assumes a convex, ellipsoidal shape, then a bi-concave, disc-like structure, and finally the shell deforms into a bowl. We note the similarity between the bi-concave discs we obtain and the shape of synthetic red blood cells \cite{doshi}, despite the differences in their respective physical origins and sizes. The rightmost column shows the results for the most rigid membrane under study. Due to the high energy penalty associated with bending, the shell remains spherical even at $c = 0.1$ M. However, upon further lowering of $c$, we first witness an ellipsoidal shape and then a flattened disc-like structure at $c = 0.005$ M. 
It is worth noting that the discs and bowls we obtain, closely resemble the shapes of elastic structures in Ref.~\cite{hamada} that are synthesized using light as a tool to engineer shape.

\begin{figure}
\centerline{
\includegraphics[scale=0.1]{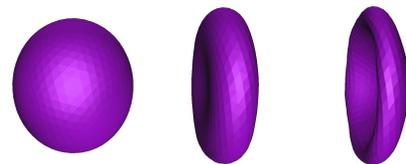}
}
\caption
{
\label{fig2}
Shell shapes that minimize free energy $\mathcal{F}$ for fixed $\kappa = 5$ and $c = 0.015$ M as a function of increasing $z =$ 0.3, 0.6, 1 (from left to right). As $z$ increases, the strength of the electrostatic interactions increase and the shell transforms from a convex, ellipsoidal form to a dimpled disk and finally to a concave bowl-like structure. All shapes correspond to the same total volume. See text for the meaning of symbols.
}
\end{figure}
Next, we study the effects of modulating the strength of Coulomb interactions on shell shape. 
In Fig.~\ref{fig2} we show snapshots of minimum-energy shell conformations when we vary the parameter $z$ keeping the flexibility of the shell and the salt concentration in the environment constant ( $\kappa = 5$ and $c = 0.015$ M). 
Changing $z$ corresponds to simulating shells with different total charge on the surface.  
The shapes from left to right correspond to the values of $z =$ 0.3, 0.6, 1. We find that at $z=0.3$, the shell assumes a convex ellipsoidal shape. As $z$ is increased to 0.6, the ellipsoid deforms into a dimpled disk and finally at $z = 1$, the bowl structure is obtained.

\begin{figure}
\centerline{
\includegraphics[scale=1]{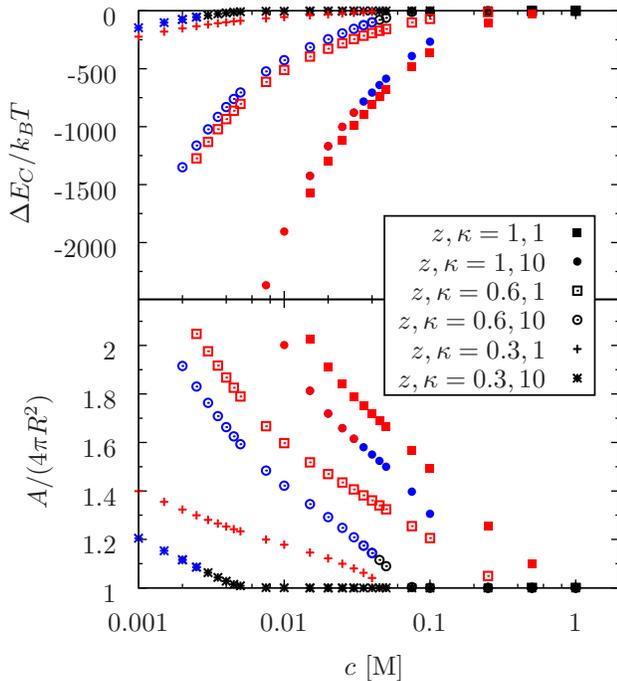}
}
\caption
{
\label{fig3}
Electrostatic contribution to the energy of the shell (Top plot) and shell's surface area (Bottom plot) vs salt concentration $c$ for different lowest-energy structures. We plot the electrostatic energy, $\Delta E_C = E_C - E_{C,S}$, which is measured  relative to that of a spherical shell with identical parameters. Similarly, the area $A$ of the shell is normalized by the area of a sphere with the same volume. Black symbols are spheres or ellipsoids; blue symbols are discs; red symbols are bowl-shaped structures. The inset shows the legend for the symbols used in the plot. The large (negative) changes in Coulomb energy help drive the shape transitions. 
}
\end{figure}

The transition to nonspherical shapes is accompanied by a decrease in the electrostatic energy. In Fig.~\ref{fig3} (top half), we plot $\Delta E_C$, the total Coulomb energy of the final structure relative to that of the spherical shell with identical parameters. The data for $\Delta E_C$ is shown as a function of $c$ for various values of $z = $ 0.3, 0.6, 1 and $\kappa = $ 1, 10. In all cases, $\Delta E_C$ is negative. For convex shapes (spheres and ellipsoids), represented by black symbols, $\Delta E_C$ is small. On the other hand, for discs and bowls, represented by blue and red symbols respectively, the reduction in electrostatic energy is more pronounced. In general, as the concentration $c$ is lowered, the behavior of $\Delta E_C$ suggests that the spherical shell deforms to an ellipsoid, then to a disk and finally to a bowl. We find that the nonspherical shapes have a larger surface area relative to the spherical conformation (see Fig.~\ref{fig3}, bottom half). We expect this to be the case as for a given fixed volume, sphere has the lowest surface area. 
We find in some cases, the minimum-energy structure has twice the surface area of a sphere with same volume.
Though a more general model of the elastic shell would include an energy penalty associated with increasing the surface area, we expect the shape changes to occur in situations where the surface energy increase due to the rise in area is compensated by the adsorption of molecules (such as neutral surfactants) to the membrane, thereby reducing its surface tension. Using the data in Fig.~\ref{fig3} we estimate that the shell surface tension should be low, $\mathcal{O}(1)$ dyne $\textrm{cm}^{-1}$, for the aforementioned predicted shapes to be realized. 

In Fig.~\ref{fig4}, we show the distribution of local electrostatic and elastic energies on the disc (top two rows) and bowl (bottom two rows). The disc corresponds to the case of $z = 0.6$, $\kappa = 10$, $c = 0.005$ M and the bowl shape is characterized by $z = 0.6$, $\kappa = 1$, $c = 0.1$ M. The electrostatic energy at a vertex is computed by summing over the screened Coulomb interactions of the charge at that vertex with all other charges on the shell. As the scalebars on the right point out, the electrostatic energy is the dominant of the two energies and drives the shape formation, with the elastic energy adapting locally to conform to the new shape. For both disc and bowl, the local elastic energy (second and fourth row) has large spatial variations and tends to be higher on the more bent regions of the nanoshell. For the disc shape, the Coulomb energy (first row) is higher near 
the center. This is, in part, due to the enhanced repulsion resulting from the proximity of the opposite faces which are at a distance less than the Debye length associated with this system. 

\section{Discussion}
Increasing the range or strength of electrostatic interactions enhances the Coulomb repulsion between any two charged vertices, making them move apart. However, the resulting extension in edge lengths is penalized by the rise in the stretching energy. In addition, the bending energy term penalizes any sharp changes in curvature, thus favoring transitions to smooth shapes. This competition between the electrostatic and elastic energies sets an effective area for the nanomembrane which in conjunction with the fixed-volume constraint determines the eventual shape of the nanoshell. Varying the screening length or the total charge on the shell changes this effective area, leading to variations in the shell shape. 

\begin{figure}
\centerline{
\includegraphics[scale=0.025]{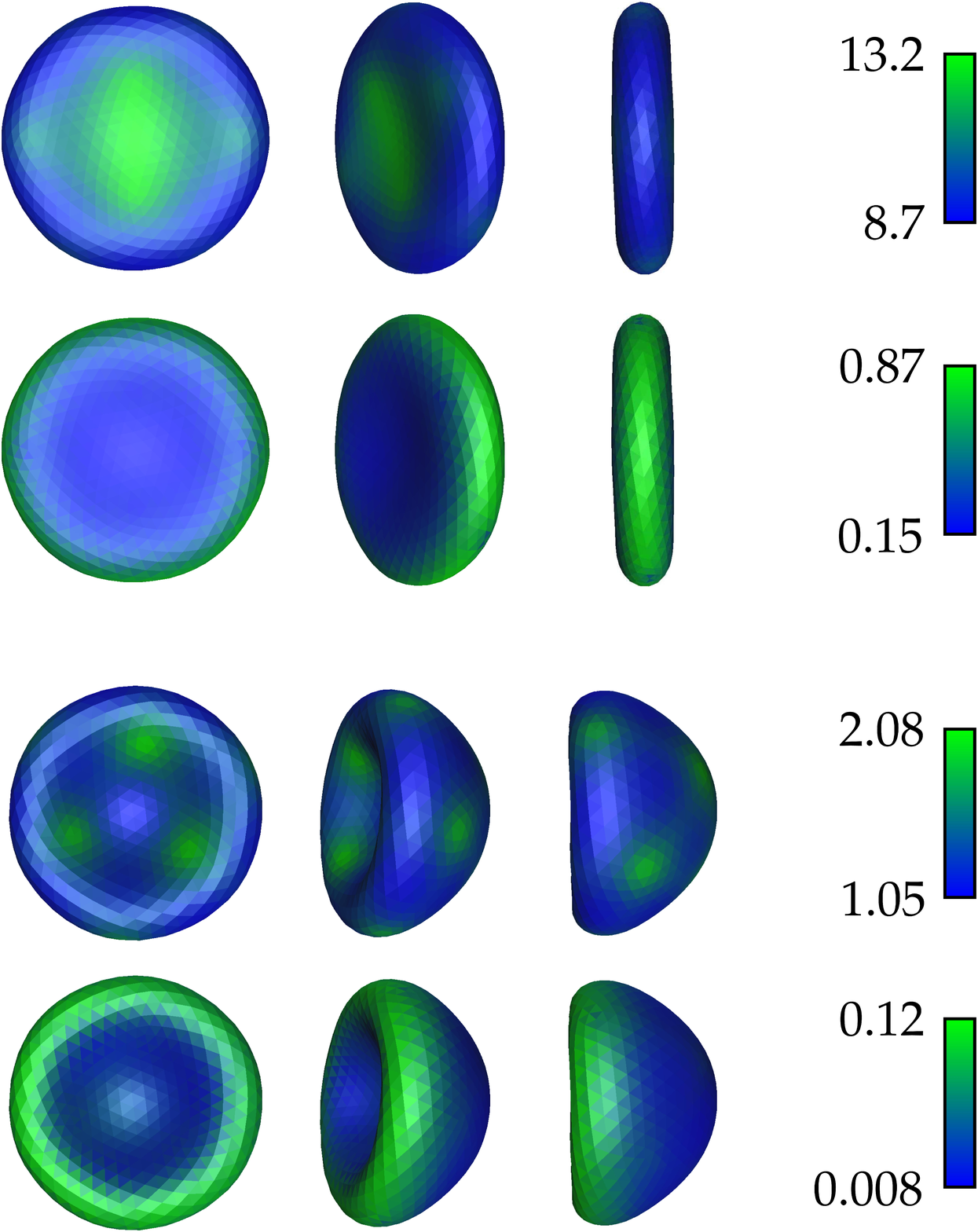}
}
\caption
{\label{fig4}
Spatial distribution of electrostatic and elastic energies (in units of $k_{\textrm{B}}T$, where $T$ is the room temperature) on the surface of the disc (top two rows) and bowl (bottom two rows). Left column: front view, middle column: angle view, right column: side view. For either shapes, the elastic energy (second and fourth row) is concentrated in the edges. The electrostatic energy on the disc (first row) is higher in the center where the opposite faces are nearby (first row). The five-coordinated vertices, which are visible as spots in the electrostatic energy distribution, lead to small fluctuations in the energy.
}
\end{figure}

To substantiate the above explanation, we focus on the sphere-to-ellipsoid-to-disc part of the observed shape transitions and perform analytical calculations. Judging by the simulation snapshots (see the images in Fig.~\ref{fig1}, rightmost column), these shapes can be approximated as oblate spheroids with different degrees of eccentricity $e$ and major semiaxis lengths $a$. Since the volume of the shell is fixed, the oblate spheroidal shell can be characterized by a single parameter $e$. For $e\to0$, one obtains sphere-like shapes and $e\to1$ leads to disc-like conformations. The competition between elastic and electrostatic energies can now be considered as determining the eccentricity $e$ for the oblate spheroid. The concentration $c$ is seen as the control over $e$ such that the lowering of $c$ can be understood as an increase in $e$. Thus, we can verify the order of shape transitions observed in our simulations by examining the change in the electrostatic energy of a 
uniformly-charged shell as its eccentricity is increased. 

For simplicity, we consider unscreened Coulomb interactions in the following calculations. We evaluate the electrostatic energy $U$ of a uniformly-charged oblate spheroidal shell with total surface charge $zN$ and with volume constrained to $\Omega = (4/3)\pi R^3$ (derivation in \emph{SI Text}). We obtain:
\begin{equation}\label{eq:UVfix}
\begin{split}
&U(e,z,N) = l_B\frac{z^2N^2}{2R}\frac{ie(1-e^2)^{1/6}}{\left(e + (1 - e^2) \textrm{tanh}^{-1} e\right)^2} \times \\
&\sum_{n \in \textrm{even}} (2n+1) P_{n}\left(i \frac{\sqrt{1-e^2}}{e}\right)  Q_n\left(i \frac{\sqrt{1-e^2}}{e}\right) \left(I_{n}(e)\right)^2,
\end{split}
\end{equation}
where $n$ is an even integer, $P_n$ and $Q_n$ are Legendre polynomials of first and second kind,
$I_{n}(e) = \int_{0}^{\pi} \sqrt{1 - e^{2} \textrm{sin}^{2} v} \,P_n(\textrm{cos}v) \textrm{sin}v \, dv$, and $U$ is evaluated relative to the thermal energy at room temperature. We define $dU$ as the electrostatic
energy of the oblate spheroid relative to the electrostatic energy of the sphere with identical parameters.
We examine the variation of $dU$ vs $e$ for the parameter set associated with the transition recorded in the open blue circles of Fig. 3 and find that the Coulomb energy of an oblate spheroidal shell subject to the constraint of constant volume decreases with increasing its eccentricity (see Fig. S2). 
In other words, a disc-shaped shell has lower Coulomb energy than a sphere of the same volume. The order of shape transitions observed in our simulations is thus backed by the above analytical result. Next, we examine the spatial distribution of the local electrostatic energy $U_{\textrm{L}}$ on the surface of the shell (see \emph{SI Text} for details). We find that for a spherical shell, $U_{\textrm{L}}$ is constant everywhere. However, as the eccentricity increases, the surface distribution of electrostatic energy becomes increasingly 
inhomogeneous. In particular, for $e=0.95$, which corresponds to a disc-like shape, we find $U_{\textrm{L}}$ varies significantly on the disc surface, assuming higher values near the disc center and low magnitude near the edge of the disc (see Fig. S3). It is evident from the top row of Fig.~\ref{fig4} that we observe this trend in our simulation results as well.

We obtain more insight into our results by exploring the low-energy conformations of a very flexible uniformly-charged shell where the elastic energy can be neglected in comparison with the Coulomb energy. Equilibrium shapes of such a shell will correspond to the minimum of the total electrostatic energy. In Eq.~\eqref{eq:UVfix}, taking the limit $e\to 1$ gives $U=0$, which is the lowest possible value for the Coulomb energy of a uniformly-charged shell. This limit corresponds to a disc-like spheroidal shell whose area approaches infinity. Further, we check that when the enclosed volume is held fixed, the Coulomb energy of a prolate spheroidal shell vanishes as well when the shell is stretched into a long and thin wire-like shape. Thus, we obtain (at least) two distinct shell shapes that correspond to the state of lowest electrostatic energy. This result suggests that in our original model system, electrostatic interactions drive the transformation 
in the shell shape by favoring the deformation of sphere towards disc-like shapes, while the elastic energies compete with the Coulomb energy to generate oblate-shaped (ellipsoidal, disc-like) structures of various eccentricities. It also appears that the elastic energy component of the free energy favors the formation of oblate shapes to prolate ones.

The constraint of fixed enclosed volume is critical to the low-energy shell conformations obtained in our simulations. If instead of the volume, the shell surface area is fixed, we expect the gallery of lowest free-energy conformations to look different than Fig.~\ref{fig1}. We check that under the constraint of fixed area, the Coulomb energy of an oblate spheroidal shell is higher when its eccentricity increases, and the spherical shape corresponds to the conformation with the lowest Coulomb energy among all oblate shapes. 
However, sphere is \emph{not} the configuration that minimizes the shell electrostatic energy when prolate-shaped deformations are considered. We find that prolate spheroids of high eccentricities have lower Coulomb energy than the sphere and the lowest-energy conformation for the area-constrained system is a prolate spheroidal shell with its major-axis length stretched to infinity. 
Hence, for the area-constrained problem, we expect the competition between Coulomb and elastic energies to give rise to different nonspherical shapes as ground-state solutions. 

In our charged shell model, we assume that the counterions remain in the bulk and do not condense on the shell surface. However, in an experiment it is possible that a fraction of the counterions do condense and in that event it becomes important to analyze their effect on the observed shape transitions. We measure this effect qualitatively in the salt-free limit for the sphere-disc transition by employing the expression for the electrostatic energy of a uniformly-charged oblate shell in a two-state model of free and condensed counterions \cite{alexander,monica}. We consider a spherically-shaped Wigner-Seitz (WS) cell of volume $V_{\textrm{WS}}$ with a single shell of volume $\Omega=(4/3)\pi R^3$ and surface charge $zN$ placed at its center. We define the quantity $\eta = \Omega / V_{\textrm{WS}}$ as the shell volume fraction. The cell also contains $N$ counterions, each of charge $-z$ to neutralize the shell charge. We separate the counterions into two distinct groups: free ions and condensed ions. Free ions occupy the available space in the WS cell which in the dilute limit becomes the volume of the cell. The condensed counterions are restricted to have translational motion in a thin layer of volume $V = \mathscr{A}(e,R) b$ surrounding the shell, where $\mathscr{A}(e,R)$ is the surface area of the oblate shell and $b=1/(2\pi l_{\textrm{B}}\sigma)$ is the Gouy-Chapman length that is chosen as the condensed-layer width. Here, $\sigma$ is the unrenormalized surface charge density. When a shell has a higher $\sigma$ or the system is characterized by a longer $l_{\textrm{B}}$, we expect the condensed-layer width to shrink owing to the enhanced counterion-shell attraction. Our choice of $b$ as the layer thickness correctly reflects this behavior. 
As $b$ is a characteristic of the charged planar surface, our analysis is limited to the regime where $b$ is much smaller than the lengths of the major and minor semiaxis of the shell. 

We write the free energy (in units of $k_BT$) associated with the shell in the event of ion condensation as:
\begin{eqnarray}\label{eq:F}
F(\alpha,e) &=& U(e,z,(1-\alpha)N) + 
\alpha N \, \textrm{ln} \left(\frac{\alpha N \Lambda^3}{\mathscr{A}(e,R)b}\right) - \alpha N \nonumber \\ 
&& +\, (1-\alpha)N \, \textrm{ln} \left( \frac{(1-\alpha)N\Lambda^3}{V_{\textrm{WS}}} \right)
- (1-\alpha)N
\end{eqnarray}
where $\alpha$ is the fraction of counterions that condense and $\Lambda$ is the thermal de Broglie wavelength. Here, the first term is the electrostatic energy of the shell obtained from Eq.~\eqref{eq:UVfix} by replacing $zN$ with the reduced charge $z(1-\alpha)N$, the next two terms stem from the entropic contribution of the $\alpha N$ condensed ions, and the last two terms correspond to the entropy of $(1-\alpha) N$ free counterions. $F$ can be considered as a function of two variables: eccentricity $e$, which characterizes the shape of the shell, and condensate fraction $\alpha$, which measures the renormalized charge on the shell. For a given $e$, we find the condensate fraction that extremizes the above free energy $F$. Using $\alpha$, we evaluate the equilibrium free energy difference, $dF$, between the free energy of the oblate shell and that of the sphere of the same volume (see \emph{SI Text} for details).
We compute $dF$ for the parameters associated with the transition recorded in the open blue circles of Fig.~\ref{fig3} and find that for all values of the volume fraction $\eta$, $dF$ becomes increasingly more negative as the eccentricity $e$ is raised, implying that the shape transitions from sphere to oblate spheroids are favored (see Fig. S4). Additionally, we find that the condensate fraction $\alpha$ decreases with increasing $e$ for all values of $\eta$. For low $\eta$, we obtain $\alpha\sim 0.1$ while for large $\eta$, we find the condensate fraction to be $\alpha\sim 0.5$. Regardless of the amount of condensation, we find the shell with higher eccentricity is preferred energetically.

We next examine the variation of the renormalized electrostatic energy $dU$ with $e$ for different $\eta$ values. For low and high values of the volume fraction, we find that $dU$ is negative and decreases, just like $dF$, upon the increase of the eccentricity (see Fig.~S5). However, for some intermediate $\eta$ values, we observe that $dU > 0$, that is the electrostatic energy increases as $e$ is raised, in sharp contrast to the free energy associated with the shell. This suggests that for some values of shell volume fractions, the shape transitions are expected to occur despite an increase in the electrostatic energy. We attribute the feasibility of such transitions to the gain in entropy by the ions as less number of ions condense when the shape is deformed from a sphere to an oblate.

The main conclusions reached above remain unchanged when we repeat the two-state model analysis assuming that the shell is an equipotential surface. For all values of $\eta$, $dF$ increasingly becomes more negative as the eccentricity $e$ is raised, implying that the shape transitions from sphere to oblate spheroids are favored (see Fig.~S6). 
Thus, judging by the variation of $dF$ determined by the above two-state model analysis, we conclude that the shape transitions from sphere to oblates of increasing eccentricity should be feasible in the event of ion condensation. However, due to the renormalization of the charge on the shell surface, it is likely that the specific parameter values (for example, concentration strength, bending rigidity) for which the shape transitions occur, will change. Quantitative results can be obtained by including counterions explicitly in the simulations and taking into account the induced polarization charges on the shell surface in analyzing changes in shape. We note that our MD-based simulation algorithm provides an ideal platform to include these effects via its coupling with recently introduced energy-functional-based approaches of treating dielectric inhomogeneities \cite{jso1,jso2}.

\section{Conclusion}
We investigate the prospects of electrostatics-based generation and control of shapes in materials at the nanoscale. We find that by increasing the strength or the range of Coulomb interacting potential, a uniformly-charged spherical shell, constrained to maintain its volume, deforms to structures of lower symmetry, resulting in ellipsoids, discs, and bowls. This symmetry breaking is accompanied with a reduction in the overall electrostatic energy of the shell and a significant spatial variation in the local elastic energy on the shell surface. To support our simulation findings, we show analytically that a uniformly-charged disc-like spheroidal shell has a lower Coulomb energy than a spherical shell of the same volume. In order to evaluate the renormalization of shell charge due to non-linear effects, we use a two-state model of free and condensed ions. We find that the shape transitions are feasible in the event of ion condensation for a wide range of shell volume fractions.

Shape changes in our model membrane are triggered by changing the attributes of the environment external to the membrane, such as the electrolyte concentration in the surrounding solvent. 
This is in contrast with transitions brought about by patterning the shell surface with defects \cite{mahadevan,lidmar,siber} or by introducing elastic inhomogeneities on the shell surface  \cite{vernizzi2,datta,rastko2,funkhouser}. In comparison with ionic shells \cite{vernizzi,rastko1}, where the primary experimental challenge is to synthesize membranes with desired stoichiometric ratios \cite{rastko1}, our base shell surface is uniformly charged and elastically homogeneous, which is relatively simple to design. 

We envision that the electrostatics-driven shell design mechanism proposed here can function as a useful template for synthesizing nanoparticle-based drug delivery carriers of desired shapes \cite{discher1,mitragotri1,loverde}. Our results can also prove useful in the analysis of shape changes in charged emulsions or reverse micelle systems that form during the metal-extraction processes involved in the recovery of scarce rare-earth elements or cleaning of nuclear waste \cite{hoogerstraete}. In addition, our findings can aid in the development of 
theories explaining the properties of stretchable electronic materials such as dielectric elastomers where electrostatic field and deformation are intimately coupled \cite{suo,whitesides}.



\subsection{Molecular Dynamics Simulation of Charged Elastic Shells}\label{sec:modelmethod}
We generate the triangulation on the shell via the Caspar and Klug construction \cite{caspur}, which produces a lattice where each point has six neighbors with the exception of 12 five-coordinated vertices (defects). Due to the presence of these defects, the lattice has a non-vanishing initial stretching energy. We remove this residual strain by appropriately choosing rest lengths of the edges \cite{katifori}, leading to a vanishing stretching energy for the initial mesh. The defects, however, lead to slight variations in the surface charge density and local elastic and electrostatic energies. By choosing a large number of lattice points, the effect of these small deviations on the resulting shape transformations is minimized. In order to make sure that our results are independent of the particular triangulation, we perform simulations employing sufficient number of lattice points generated via different choices of Caspar and Klug constructions, obtaining similar 
results for all runs.   
 
Our lattice maintains the initial connectivity throughout the shape evolution. Since each vertex carries a charge of the same sign, our discretized membrane is characterized with an inbuilt self-avoidance due to the mutual electrostatic repulsion between any pair of vertices. However, to ensure complete stability, we include an additional short-range, purely repulsive Lennard-Jones potential between two vertices, where each vertex is modeled as a hard sphere with radius chosen to be a fraction of the average edge length $a$ associated with the triangular lattice.

We use molecular dynamics (MD) method to minimize the free energy $\mathscr{F}$ which requires the analytical expressions for the gradients of $\mathscr{F}$ with respect to the vertex positions. Evaluating the gradient of the bending energy term is relatively difficult and we show this calculation in the \emph{SI Text}. Our simulations start from a spherical shell with a nearly homogeneous surface distribution of local elastic and electrostatic energies. Slight deviations in the energies arise from the presence of five-coordinated vertices which are the result of using the aforementioned triangulation of the shell surface. We assign the vertices a kinetic energy $\mathcal{K} = \sum_{i}(1/2) \mu \dot{\mathbf{r}}_{i}^{2}$, and direct their motion according to the forces derived from $\mathcal{F}$, where the latter plays the role of the potential energy. We thus obtain the Lagrangian $L = \mathcal{K}  - \mathcal{F}$, from which we derive the equations of motion for the vertices: $\mu \ddot{\mathbf{r}
}_{i} = -\nabla_{\mathbf{r}_{i}} \mathcal{F}$. Here, $\mu$ is a mass term associated with the vertices which determines the choice of the simulation timestep. These equations of motion, which form the basis of the MD simulation of the vertices, are appropriately augmented to preserve the constraint of fixed total volume. We achieve this via the Shake-Rattle routine \cite{shake} of implementing constraints that guarantees the conservation of shell volume at each simulation step. Finally, in order to arrive at the shape that corresponds to the minimum of the energy landscape we couple the MD scheme with simulated annealing. 
We associate a (fictitious) temperature with the kinetic energy of the vertices and employ a Nose-Hoover thermostat to set it. This temperature is not the physical temperature, it is merely a parameter we employ to control the annealing process. We reduce this temperature at periodic
intervals so as to arrive at the lowest point of the potential energy associated with the MD Lagrangian, thus reaching the minimum of the free energy $\mathscr{F}$.

\begin{acknowledgments}
We thank R. Sknepnek, Z. Yao, G. Vernizzi, and J. Zwanikken for many insightful discussions. 
The model was developed with the financial support of the Office of Basic Energy Sciences within the Department of Energy (DOE) grant number DE-FG02-08ER46539. 
The computational work was funded by the Office of the Director of Defense Research and Engineering (DDR$\&$E) and the Air Force Office of Scientific Research (AFOSR) under Award No. FA9550-10-1-0167.
\end{acknowledgments}


\end{article}

\end{document}